\documentclass[12pt]{article}
\usepackage{amsmath,epsfig,graphicx,amssymb}
\usepackage[margin=0.7in]{geometry}
\usepackage{float}
\date{}
\begin{document}
\title{Relationship between classical and quantum mechanics in micellar aqueous solutions of surfactants}
\author{Partha Ghose\footnote{partha.ghose@gmail.com}\\Tagore Centre for Natural Sciences,  Philosophy,\\ Rabindra Tirtha, New Town, Kolkata 700156, India,\\
and\\
Yuri Mirgorod, Southwest State University, Russia, 305040, Kursk.}
\maketitle

\begin{abstract}
Micellar aqueous solutions of ionic surfactants have been observed to exhibit proton delocalization (the nuclear quantum effect) and to oscillate between a low density (LDL) and a high density (HDL) state of water state at a fixed temperature. It is shown in this paper that such phenomena can be explained with the help of the interpolating Schrödinger equation proposed by Ghose (Ghose, 2002). The nuclear quantum effect can be described by the tunneling of a harmonic oscillator in a symmetric double-well potential, and an ensemble of harmonic oscillators can model the LDL-HDL oscillations. The thermodynamics of such harmonic oscillators has been worked out showing continuous transitions between the quantum and classical limits.
\end{abstract}

\section{Introduction}
Classical mechanics is a fully developed field whereas quantum mechanics is still a developing field. The interplay of classical and quantum mechanics was noticed for the first time by Planck while studying the radiation spectrum of a completely black body, and it was Planck who is considered as the originator of quantum theory. Planck's formula is in excellent agreement with experimental data on the distribution of energy in the emission spectra of a black body over the entire range of frequencies and temperatures and is used in pyrometry. The ammonia molecule consists of three hydrogen atoms located in the plane of an equilateral triangle with a nitrogen atom at the top. The pyramid inverts when the nitrogen atom changes position from one equilibrium point to another due to the tunnel effect. This inversion is repeatedly switched at a tunneling rate of 2.4×1012 Hz, which agrees well with the calculated value by solving the Schr\"{o}dinger equation in a double-well potential. This quantum mechanical effect is applied in the maser.

{{\flushleft{\em Chemistry of micelle formation}}}
\vskip0.1in

Consider a micellar aqueous solution system. Micelle formation is accompanied by a liquid-liquid phase transition in an ensemble of small (nanometer size) water systems. Near the critical micelle concentration or the midpoint of the liquid-liquid phase transition, the electron densities of thermal and quantum fluctuations come to the fore \cite{mir}. 

In classical mechanics thermal fluctuations occur when the oscillator is coupled to a heat bath of temperature $T$. At finite temperature quantum statistical mechanics allows the description of the transition from pure quantum fluctuations at $T = 0$ to classical thermal fluctuations in the high temperature limit. The energies of thermal and quantum fluctuations of water correlate and compete. If $\hbar \omega \ll k_BT$, thermal fluctuations prevail; if $\hbar\omega \gg k_BT$, then quantum fluctuations prevail, and in case $\hbar\omega \approx k_BT$ the quantum and thermal fluctuation are approximately equal. The thermodynamics and statistical mechanics of the harmonic oscillator with and without zero-point energy (ZPE) have been studied by Boyer \cite{boy} and Pe$\tilde{n}$a et al \cite{pen}. Mirgorod and Storozhenko \cite{mir2} have studied the role of ZPE of water in micelle formation of ionic surfactants. They find that a `competition of thermal and 
quantum fluctuations depends on temperature, the volume of amphiphilic molecules, and additives'. 

It has been observed that room-temperature water is not simply a molecular liquid---its protons experience wild excursions (delocalization) along the hydrogen bond (HB) network driven by quantum fluctuations, which result in an unexpectedly large probability of transient autoionization events. This is known as the `nuclear quantum effect'. 

A team at the Southwest State University, Kursk, \cite{mir2} discovered that the equation for entropy-enthalpy compensation in the process of micellization of ionic, non-ionic surfactants, and dissolution of hydrocarbons is related to the equation of the total energy of a quantum harmonic oscillator
\begin{equation}
\Delta H_i = T_c\Delta S_i + \Delta H_{T=0}\label{comp} 
\end{equation}
where $T_c$ is the iso-equilibrium temperature and $\Delta H_{T=0}$ is a constant. There is still no theoretical explanation of the constant. In fact, in Ref. \cite{liu} it is even considered mystical. It has been shown empirically that this constant is the zero-point energy (ZPE) of the O-H harmonic oscillator. Therefore, the S/H compensation equation (\ref{comp}) can be witten as
\begin{equation}
\Delta H_i = T_c\Delta S_i + \frac{1}{2}\hbar\omega
\end{equation}
where $\omega$ is the angular frequency of the O-H oscillator which can be determined from the fundamental modes of the IR spectrum of water at room temperature. This equation is similar to the well known equation for the average energy of a quantum harmonic oscillator
\begin{equation}
\langle \epsilon_{qm}\rangle = \hbar\omega\left(\langle n\rangle + \frac{1}{2}\right) 
\end{equation}
Now, according to Sch\"{o}nhammer the non-zero ground-state implies quantum fluctuations around the minimum of the potential with the mean square value proportional to Planck's constant \cite{schon}. 

All these effects can be described by the effective Hamiltonian 
\begin{eqnarray}
H^{eff}_{\lambda}= \begin{pmatrix}
\Delta \text{H}^{(\lambda)} + \frac{1}{2}\hbar\omega^{(\lambda)} & (1-\lambda)\frac{1}{2}\Delta^{(\lambda)}\\
(1-\lambda)\frac{1}{2}\Delta^{(\lambda)} & \Delta \text{H}^{(\lambda)} +\frac{1}{2}\hbar\omega^{(\lambda)}
\end{pmatrix}\label{Ham1}
\end{eqnarray}
where $\Delta \text{H}^{(\lambda)}$ is the enthalpy change in micelle formation, $\frac{1}{2}\hbar\omega^{(\lambda)}$ is the ZPE of water, and the off-diagonal terms are responsible for tunneling (proton delocalization). The parameter $\lambda$ varies with the temperature, volume of amphiphilic molecules as well as on additives \cite{mir}. As $\lambda \rightarrow 0$, the system becomes more and more isolated and more and more quantum mechanical, whereas as $\lambda \rightarrow 1$, it becomes more and more coupled to its  environment and more and more classical.

{{\flushleft{\em Background for the harmonic oscillator model of micelle formation}}}
\vskip0.1in

We will show in what follows that this effective Hamiltonian can be derived from an interpolating Schr\"{o}dinger equation proposed by Ghose \cite{gh1, gh2}. We will first show that this interpolating equation has some remarkable consequences for a particle in a box and a particle in a symmetric double-well potential. We will then proceed to use the equation to study the thermodynamics of a harmonic oscillator.

One of the fundamental problems in conceptually connecting quantum mechanics and classical mechanics is that in the former physical states are described by complex square integrable functions spanning a Hilbert space (or generally by vectors in linear vector spaces known as Hilbert spaces). In classical mechanics points in phase space (the six-dimensional space spanned by the position and momentum coordinates) describe states. The first attempt at bridging this conceptual gap was taken by Koopman \cite{koop} and von Neumann \cite{vn} who formulated the classical mechanics of a system by representing its states by complex square integrals functions $\psi(x,p)$ of the space and momentum coordinates spanning a Hilbert space. These `wave functions' are postulated to satisfy the Liouville equation, and as a result, the system is described by a conserved density function $\rho(x,p)$ in phase space, as in classical statistical mechanics. This guarantees that there are no interference effects. Consequently, however, the transition to quantum mechanics, in which interference effects play a fundamental role, remained problematic \cite{mauro}.

The next step was taken by Ghose \cite{gh1, gh2} who wrote down an interpolating Schrödinger-like equation rather than a Liouville equation for an interpolating wave function with an additional term $\lambda Q$, where $Q$ is the `quantum potential' and $\lambda$ an arbitrary parameter which can vary between 0 and 1. In the limit $\lambda =1$ all quantum mechanical effects are eliminated and the system is fully classical, whereas in the limit $\lambda=0$ one recovers the Schr\"{o}dinger equation and full quantum mechanical coherence is restored.

\section{Interpolating Schr\"{o}dinger Equation}

Consider first the classical Hamilton-Jacobi equation \cite{gold}
\begin{eqnarray} 
\frac{\partial S(\vec{x}, t; \vec{\alpha})}{\partial t} + H(\vec{x}, \nabla S(\vec{x}, t; \vec{\alpha}), t) &=& 0,\nonumber\\
H(\vec{x}, \nabla S(\vec{x}, t; \vec{\alpha}), t) &=& \frac{(\nabla S(\vec{x}, t; \vec{\alpha}))^2}{2m} + V(\vec{x},t),\label{1}
\end{eqnarray}
and the continuity equation
\begin{eqnarray}
\frac{\partial \rho(\vec{x}, t; \vec{\alpha})}{\partial t} + \vec{\nabla}.(\vec{v}\rho(\vec{x}, t; \vec{\alpha})) = 0\label{2}
\end{eqnarray}
for the probability density $\rho(\vec{x}, t; \vec{\alpha})$ in phase space. The particle velocity is defined by
\begin{equation}
\vec{v}(\vec{x}, t; \vec{\alpha}) = \frac{d(\vec{x}, t; \vec{\alpha})}{dt} = \frac{1}{m}\nabla S(\vec{x}, t; \vec{\alpha})
\end{equation}
and integration w.r.t. $t$ gives the trajectories $\vec{x}(t; \vec{\alpha})$.
The parameter $\vec{\alpha}$, an integration constant, indicates that each particle has its own velocity at every point$(\vec{x}, t)$. These constitute the fundamental equations of classical statistical mechanics. 

Eqns (\ref{1}) and (\ref{2}) can be combined into a single classical Schr\"{o}dinger equation  
\begin{eqnarray}  
i\,\hbar\,\frac{\partial \psi_{cl}(\vec{x}, t; \vec{\alpha})}{\partial t} &=& \left[\hat{H}_{qm} - Q_{cl}\right]\psi_{cl}(\vec{x}, t; \vec{\alpha}) \equiv \hat{H}_{cl}\psi_{cl}(\vec{x}, t; \vec{\alpha}), \label{Sch2}\\
Q_{cl} &=& -\frac{\hbar^2}{2 m}\,\frac{\nabla^2 \sqrt{\rho(\vec{x}, t; \vec{\alpha})}}{\sqrt{\rho(\vec{x}, t; \vec{\alpha})}}%
\end{eqnarray} 
for a complex classical wave function 
\begin{equation}
\psi_{cl}(\vec{x}, t; \vec{\alpha}) = \sqrt{\rho(\vec{x}, t; \vec{\alpha})}\,{\rm exp}(iS(\vec{x}, t; \vec{\alpha})/\hbar).\label{clwave}
\end{equation}
Substitution of this polar form of the wave function into eqn (\ref{Sch2}) and separation of the real and imaginary parts at once leads back to eqns (\ref{1}) and (\ref{2}). 
Although these equations involve the reduced Planck constant $\hbar$, it drops out of the classical equations (\ref{1}) and (\ref{2}). Hence, put in this Schr\"{o}dinger-like form, {\em classical mechanics does not require $\hbar$ to be small compared to $S$ or to vanish}.

Now consider the equation \cite{gh1, gh2}
\begin{eqnarray}  
i\,\hbar\,\frac{\partial \psi(\vec{x},t;\lambda\vec{\alpha})}{\partial t} &=& \left[\hat{H}_{qm} - \lambda Q^{(\lambda)}\right]\psi(\vec{x},t;\lambda\vec{\alpha}), \label{int}\\
Q^{(\lambda)} &=& -\frac{\hbar^2}{2 m}\,\frac{\nabla^2 \sqrt{\rho(\vec{x},t;\lambda\vec{\alpha})}}{\sqrt{\rho(\vec{x},t;\lambda\vec{\alpha})}}\nonumber
\end{eqnarray}
for a complex interpolating wave function 
\begin{equation}
\psi(\vec{x},t;\lambda\vec{\alpha}) = \sqrt{\rho(\vec{x},t;\lambda\vec{\alpha})}\,{\rm exp}(iS(\vec{x},t;\lambda\vec{\alpha})/\hbar),\label{intpolwave}
\end{equation}
where $\lambda$ is a real parameter which varies continuously between 0 and 1: $0\leq \lambda \leq 1$. In the limit $\lambda = 0$ the equation reduces to the Schr\"{o}dinger equation and the wave function becomes single-valued with the momentum determined solely by the position $\vec{x}$. However, in the limit $\lambda = 1$ the equation reduces to the classical equation (\ref{Sch2}) with a multi-valued wave function and crossing trajectories. Hence, equation (\ref{int}) interpolates smoothly between quantum and classical mechanics. 

It is to be noted that eqn (\ref{int}) with $\lambda = 1$ and a wave function $\psi_{cl}(\vec{x})$ is known in the literature as the `classical Schr\"{o}dinger equation'. It implies non-crossing  trajectories as shown in Ref. \cite{ben}, and hence does not represent true classical mechanics. Hence, to obtain the correct classical trajectories which can cross, it is essential to make the classical wave function dependent on $\lambda\vec{\alpha}$ as in (\ref{int}). In the limit $\lambda = 0$ it reduces to the correct quantum mechanical wave function $\psi(\vec{x})$ which is single-valued. This aspect was not clearly spelt out in Refs. \cite{gh1, gh2}.

Explicit examples of the harmonic oscillator and free particle wave packets will be found in Ref. \cite{gh2, klaus1}, and the double-slit set up, the 2D harmonic oscillator and the hydrogen-like atom in Ref. \cite{klaus2}.

To simplify the notation we will henceforth write $\psi^{(\lambda)}(\vec{x},t)$ instead of $\psi(\vec{x},t;\lambda\vec{\alpha})$.

Notice that
\begin{equation}
\hat{H}_0 \sqrt{\rho^{(\lambda)}} = \frac{\hat{p}^2}{2m}\sqrt{\rho^{(\lambda)}} = -\frac{\hbar^2\nabla^2 \sqrt{\rho^{(\lambda)}}}{2 m} =  Q^{(\lambda)} \sqrt{\rho^{(\lambda)}},\,\,\,\, \hat{p} = -i\hbar \nabla, \label{phat}
\end{equation}
showing that $Q^{(\lambda)}$ is the quantum mechanical kinetic energy of the particle. 
Hence, if the wave function is real (bound states), eqn (\ref{int}) can be written as
\begin{eqnarray}  
i\,\hbar\,\frac{\partial \sqrt{\rho^{(\lambda)}(\vec{x},t)}}{\partial t} &=& \left[(1 - \lambda)\hat{H}_0 + V\right]\sqrt{\rho^{(\lambda)}(\vec{x},t)}.\label{real}
\end{eqnarray}
The factor $(1 - \lambda)$ is thus the fraction of the quantum mechanical kinetic energy carried by the particle. The free Hamiltonian is therefore renormalized by the factor $(1 - \lambda)$. This means that the mass is renormalized to $m^{(\lambda)} = m/(1 - \lambda)$. In the following sections we will show that this has some remarkable consequences for a particle in a box and a particle in a double well potential in the classical limit $\lambda = 1$.
\subsection{Particle in a Box}
The time independent form of equation (\ref{real}) is
\begin{equation}
\left[(1 - \lambda)\hat{H}_0 + V\right]\sqrt{\rho(\vec{x})} = E^{(\lambda)}\sqrt{\rho(\vec{x})}.
\end{equation}
Take the case of a particle in a box of length $L$. The solutions are
\begin{eqnarray}
\sqrt{\rho_n(x)} = \sqrt{\frac{2}{L}}{\rm sin}\left(k_n (x + \frac{L}{2})\right),\,\,\,\,k_n = \frac{n\pi}{L},\,\, -\frac{L}{2} < x < \frac{L}{2},\,\,n= 1,2,\cdots.
\end{eqnarray}
Thus, although the amplitudes $\rho_n$ are independent of $\lambda$, the energy eigenvalues are
\begin{eqnarray}
E^{(\lambda)}_n &=& \hbar \omega^{(\lambda)}_n = \frac{\hbar^2 k_n^2}{2m^{(\lambda)}} = (1 - \lambda)\frac{n^2\pi^2\hbar^2}{2mL^2} = (1 - \lambda)E_n
\end{eqnarray}
because of mass renormalization. This shows that the quantized energies $E^{(\lambda)}_n \rightarrow 0$ as $\lambda \rightarrow 1$. This result can in principle be verified by looking at the energy spectra of nanostructures.

\subsection{Double-Well Potential and Localization}
\begin{figure}[H]
\centering
\includegraphics[scale=0.5]{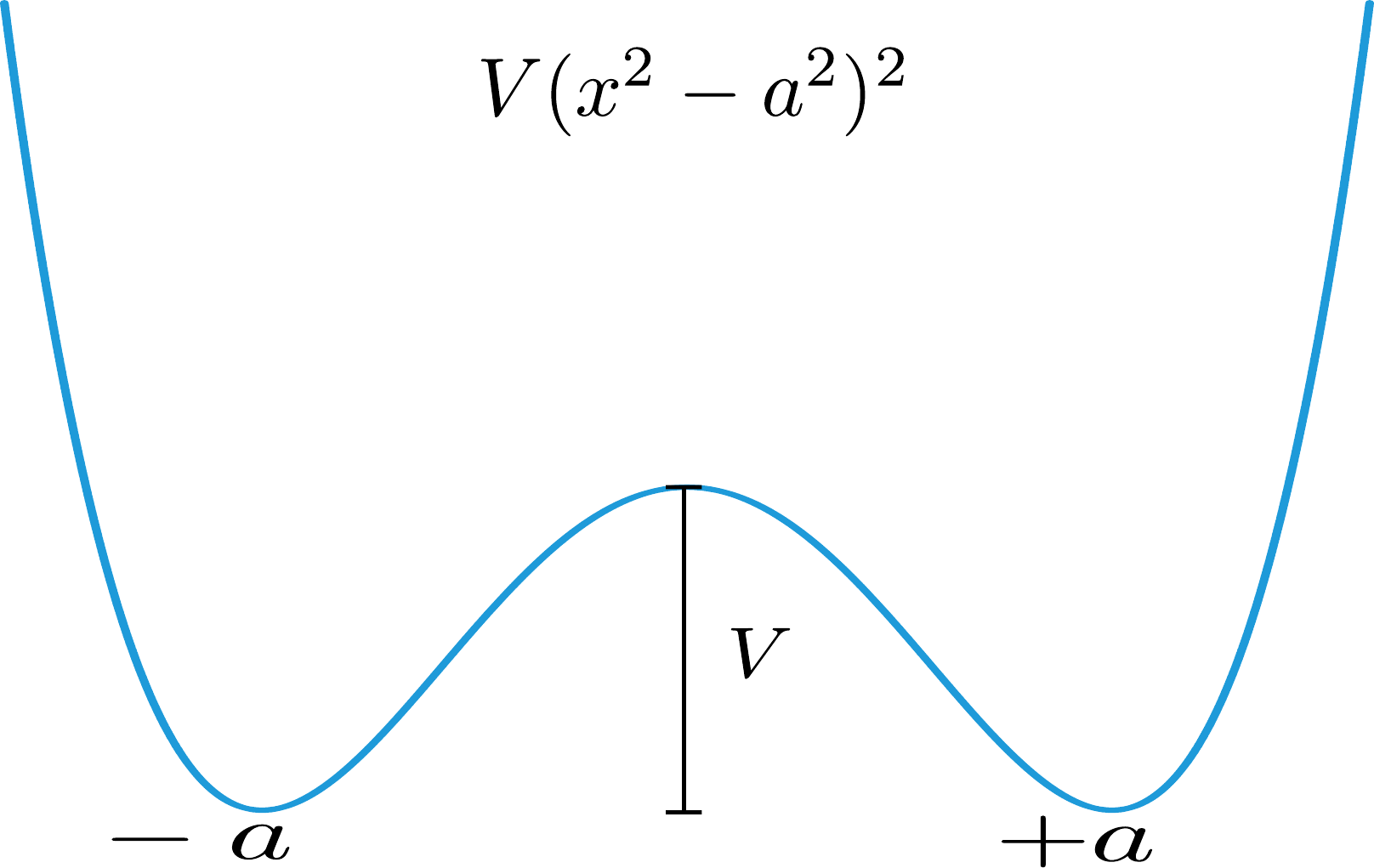}
\caption{\label{Fig. 1}{\footnotesize A double-well potential with minima (zeros) at $x = \pm a$ and of height $V_0 a^4$} at $x = 0$. Classically a particle will be trapped in one of these wells, but quantum mechanically it can tunnel through the potential barrier and oscillate between the two wells.}
\end{figure}
Eqn (\ref{real}) for a system in a double-well potential has the form
\begin{eqnarray}  
i\,\hbar\,\frac{\partial \psi^{(\lambda)}(\vec{x},t)}{\partial t} &=& \left[-(1 - \lambda)\frac{\hbar^2}{2m}\partial_x^2 + V_0(x^{2}-a^{2})^{2}\right]\psi^{(\lambda)}(\vec{x},t),\,\,\,\,\psi^{(\lambda)}(\vec{x},t) = \sqrt{\rho^{(\lambda)}(\vec{x},t)}.\label{dwe2}
\end{eqnarray}
The double-well potential has minima at $\pm a$ about which the potential form can be approximated as
\begin{eqnarray}
V_0(x^{2}-a^{2})^{2} &\approx & 4V_0a^{2}(x-a)^{2} , |(x-a)|<<a\nonumber\\
				   &\approx & 4V_0a^{2}(x+a)^{2} , |(x+a)|<<a
\end{eqnarray}
Since eqn (\ref{dwe2}) is symmetric under the interchange $x \rightarrow - x$, the solutions can be separated into symmetric and anti-symmetric superpositions of harmonic oscillator wave functions around the two minima,
\begin{eqnarray}
\psi^{(\lambda)}(x)_{(S, A)}=a_{+}\psi^{(\lambda)}_{+}(x) \pm a_{-}\psi^{(\lambda)}_{-}(x) ~,~\psi_{\pm}^{(\lambda)}(x) = \left(\frac{2\alpha^{(\lambda)}}{\pi}\right)^{1/4}e^{-\alpha^{(\lambda)} (x\mp a)^{2}}~.
\end{eqnarray}
with $|a_+|^2 + |a_-|^2 = 1$,
where the coefficient $\alpha^{(\lambda)}$ can be determined in the vicinity of these minima at $x =\pm a$: 
\begin{eqnarray}
\frac{1}{2}\hbar\omega^{(\lambda)} \psi_{\pm}^{(\lambda)}(x)&=&\left[-(1-\lambda)\frac{\hbar^{2}}{2m}\partial_{x}^{2}+4V_0 a^{2}(x\mp a)^{2}\right]\psi_{\pm}^{(\lambda)}(x)~\nonumber\\
&:=&\left[-\frac{\hbar^{2}}{2m^{(\lambda)}}\partial_{x}^{2}+\frac{1}{2}K(x\mp a)^{2}\right]\psi_{\pm}^{(\lambda)}(x)~,~m^{(\lambda)}=\frac{m}{1-\lambda}~,~K=8V_0 a^{2},\\
\omega^{(\lambda)} &=&\sqrt{\frac{K}{m^{(\lambda)}}}=\sqrt{\frac{8V_0 a^{2}(1-\lambda)}{m}}~,~\alpha^{(\lambda)} =\frac{m^{(\lambda)}\omega^{(\lambda)}}{2\hbar}=\frac{1}{\hbar}\sqrt{\frac{2V_0 a^{2}m}{1-\lambda}} \label{omega}
\end{eqnarray}
This shows that the angular frequency $\omega^{(\lambda)}$ and the coefficient $\alpha^{(\lambda)}$ are renormalized such that $\omega^{(\lambda)}$ goes to zero and the coefficient $\alpha^{(\lambda)}$, the inverse of the Gaussian width, blows up in the limit $\lambda = 1$.

The overlap between the wave functions $\psi_{+}^{(\lambda)}(x)=\sqrt{\rho_{+}^{(\lambda)}(x)}$ and $\psi_{-}^{(\lambda)}(x)=\sqrt{\rho_{-}^{(\lambda)}(x)}$ is given by
\begin{eqnarray}
P=\int dx \sqrt{\rho_{+}^{(\lambda)}(x)}\sqrt{\rho_{-}^{(\lambda)}(x)}=e^{-2\alpha^{(\lambda)} a^{2}}.
\end{eqnarray}
Now let us write the renormalized Schr\"{o}dinger equation in the basis of the approximated harmonic potential wave function,
\begin{eqnarray}
H^{(\lambda)}\psi^{(\lambda)}(x)&=&a_{+}H^{(\lambda)}\psi_{+}^{(\lambda)}(x)+a_{-}H^{(\lambda)}\psi_{-}^{(\lambda)}(x)\nonumber\\
&=&E^{(\lambda)}\left(a_{+}\psi_{+}^{(\lambda)}(x)+a_{-}\psi_{-}^{(\lambda)}(x)\right)~\label{SHO}
\end{eqnarray} 
Let us multiply eq \eqref{SHO} by $\psi_{+}^{(\lambda)}(x)$ from the right and then $\psi_{-}^{(\lambda)}(x)$ from the left and integrate over all x except $\pm a$:
\begin{eqnarray}
a_{+}\int^{a+\delta}_{a-\delta} dx \psi^{(\lambda)}_{+}(x)\left[-\frac{\hbar^{2}}{2m}(1-\lambda)\partial_{x}^{2}+4V_0 a^{2}(x-a)^{2}\right]\psi^{(\lambda)}_{+}(x)\nonumber\\ + (1-\lambda)a_{-}\int^{a+\delta}_{a-\delta}dx \psi^{(\lambda)}_{-}(x)\left[-\frac{\hbar^{2}}{2m}\partial_{x}^{2}\right ]\psi^{(\lambda)}_{+}(x) &=& E^{(\lambda)}a_{+} + a_{-}(1 - \lambda) P_{\delta}~,\\
a_{-}\int_{-a-\delta}^{-a+\delta} dx \psi^{(\lambda)}_{-}(x)\left[-\frac{\hbar^{2}}{2m}(1-\lambda)\partial_{x}^{2}+4V_0 a^{2}(x+a)^{2}\right]\psi^{(\lambda)}_{-}(x)\nonumber\\ + (1-\lambda)a_{+}\int _{-a-\delta}^{-a+\delta}dx \psi^{(\lambda)}_{-}(x)\left[-\frac{\hbar^{2}}{2m}\partial_{x}^{2}\right ]\psi^{(\lambda)}_{+}(x)&=& E^{(\lambda)}a_{-} + a_{+}(1 - \lambda)P_{\delta}~\label{eqn}
\end{eqnarray} 
where $P_{\delta}$ is defined as 
\begin{eqnarray}
P_{\delta} = \int_{-a-\delta}^{-a+\delta} \psi^{(\lambda)}_{-}(x)\left[-\frac{\hbar^{2}}{2m}\partial_{x}^{2}\right ]\psi^{(\lambda)}_{+}(x) dx =\int_{a-\delta}^{a+\delta} \psi^{(\lambda)}_{-}(x)\left[-\frac{\hbar^{2}}{2m}\partial_{x}^{2}\right ]\psi^{(\lambda)}_{+}(x) dx
\end{eqnarray}
with $\delta$ the neighborhood around the two wells. Now one can define the tunneling coefficient
\begin{eqnarray}
\frac{1}{2}\Delta^{(\lambda)} &=& -\frac{\hbar^{2}}{2m}\int_{a - \delta}^{a + \delta} \psi^{(\lambda)}_{-}(x)\partial^{2}_{x}\psi^{(\lambda)}_{+}(x) dx.  
\end{eqnarray}
Let the harmonic well energies be
$\epsilon_{\pm} = \frac{1}{2}\hbar\omega_0^{(\lambda)}$.   
Using the above equations and the integrals one obtains in the limit $\delta\to 0$ 
the Hamiltonian 
\begin{eqnarray}
H^{SHO}_{(\lambda)}= \begin{pmatrix}
\frac{1}{2}\hbar\omega_0^{(\lambda)} & (1-\lambda)\frac{1}{2}\Delta^{(\lambda)}\\
(1-\lambda)\frac{1}{2}\Delta^{(\lambda)} & \frac{1}{2}\hbar\omega_0^{(\lambda)}
\end{pmatrix}\label{Ham2}
\end{eqnarray}
in the basis of 
\begin{eqnarray}
|+a\rangle =\begin{pmatrix}
1 \\ 0
\end{pmatrix}~,~|-a\rangle =\begin{pmatrix}
0 \\ 1
\end{pmatrix}
\end{eqnarray}
From the off-diagonal elements one can see the progressive quenching of the tunneling (delocalization) and `dynamical localization' of the system near the bottom of th e wells as $\lambda \rightarrow 1$. The same result can be obtained from quantum mechanics ($\lambda = 0$) by coupling the system to an infinite bath of quantum oscillators of all frequencies annuclear quantum effectd making use of adiabatic renormalization \cite{chak1, chak2}. 

One can define a density matrix
\begin{eqnarray}
\rho = \frac{p(\lambda)}{2}I+\frac{(1-p(\lambda))}{2}[|+a\rangle +|-a\rangle][\langle +a|+\langle -a|]
\end{eqnarray}
with $p(0) = 0, p(1) = 1$.
which commutes with the Hamiltonian $H^{SHO}_{\lambda}$ because $\frac{1}{\sqrt{2}}\left[|+a\rangle +|-a\rangle\right]$ is always its eigenfunction,
\begin{eqnarray}
 \frac{1}{\sqrt{2}}\left[|+a\rangle +|-a\rangle\right] =\left[\frac{1}{2}\hbar\omega^{(\lambda)} + (1-\lambda)\frac{1}{2}\Delta^{(\lambda)}\right]\frac{1}{\sqrt{2}}\left[|+a\rangle +|-a\rangle\right]
\end{eqnarray}
leading to 
\begin{displaymath}
[H^{SHO}_{\lambda},\rho]=\frac{1}{2}[H^{SHO}_{\lambda},I]+\frac{1}{2}[H^{SHO}_{\lambda},\left[|+a\rangle +|-a\rangle\right]\left[\langle +a| +\langle -a|\right]]=0.
\end{displaymath}
This shows that $\rho$ is a mixed state in the limit $\lambda = 1$. Hence, the interpolating system is quantum mechanical with probability $\frac{1}{2}(1 - p(\lambda))$ and classical with probability $\frac{1}{2}p(\lambda)$.

The tunneling amplitude $\frac{1}{2}\Delta^{(\lambda)}$ removes the degeneracy of the states $\psi_{\pm}$ and one can write the normalized symmetric and anti-symmetric solutions with energies $(E_+, E_-)$ given by
\begin{eqnarray}
\psi_S &=& \frac{1}{\sqrt{2}}(\psi_+ + \psi_-),\,\,\,\, E_+ = \frac{1}{2} \hbar\omega_0^{(\lambda)} - \frac{1}{2}(1 - \lambda)\Delta^{(\lambda)},\\
\psi_A &=& \frac{1}{\sqrt{2}}(\psi_+ - \psi_-),\,\,\,\, E_- = \frac{1}{2} \hbar\omega_0^{(\lambda)} +\frac{1}{2}(1 - \lambda)\Delta^{(\lambda)}.
\end{eqnarray}
These states have slightly different energies, and hence the system will oscillate with a time period $2\pi\hbar/(1 - \lambda)\Delta^{(\lambda)}$. Then, provided it is in the left well at $t = 0$, it will definitely be found in the left well at all times that are integral multiples of $2\pi\hbar/(1 - \lambda)\Delta^{(\lambda)}$, and in the right well at all times that are half-odd integral multiples of $2\pi\hbar/(1 - \lambda)\Delta^{(\lambda)}$. At all other times there will be a finite probability of finding it in either well. Thus, the probability of finding it in the left well minus the probability of finding it in the right well is given by
\begin{equation}
P^{(\lambda)}(t) = \cos \left((1 - \lambda)\Delta^{(\lambda)} t/\hbar\right).
\end{equation}
The probability of finding the system anywhere between the two wells is practically zero if $V \gg \hbar\omega_0^{(\lambda)}$ ($\sim \text{exp} (-V/\hbar\omega_0^{(\lambda)})$). It should be possible to measure $P^{(\lambda)}(t)$ and determine $\Delta^{(\lambda)}$.

{\em It is evident from all these results and a comparison of eqns (\ref{Ham1}) and (\ref{Ham2}) that the latter (i.e. $H^{SHO}_{\lambda}$) is the required correction to the classical enthalpy change $\Delta H^\lambda$ in micelle formation, indicating that micelles behave like harmonic oscillators in a symmetric double-well potential. Further, these harmonic oscillators behave purely quantum mechanically when their coupling $\lambda$ to the environment is zero and non-classically when $0 < \lambda < 1$}.

\section{Thermodynamics of the Harmonic Oscillator}
In order to study the thermodyamics of the harmonic oscillator, it will be useful to use the de Broglie-Bohm theory \cite{bohm} to cast the interpolating Schr\"{o}dinger eqn (\ref{int}) in the form 
\begin{equation}
E = \frac{p^2}{2m} + V(x, a) + (1 - \lambda)Q \label{E}
\end{equation}
by writing it in the Hamilton-Jacobi form
\begin{equation}
\frac{\partial S}{\partial t} + \frac{(\nabla S)^2}{2m} + V(x, a) + (1 - \lambda)Q = 0
\end{equation}
 with the help of (\ref{intpolwave}) and using the guidance condition $p = \nabla S,\, \partial S/\partial t = -E$.
It will also be convenient to rewrite (\ref{E}) as
\begin{eqnarray}
E &=& (1 - \lambda) \left(\frac{p^2}{2m} + V(x, a) + Q\right) + \lambda \left(\frac{p^2}{2m} + V(x, a)\right)\nonumber\\  
&:=& (1 - \lambda)E_{qm} + \lambda E_{cl}.
\end{eqnarray}
Now,
\begin{eqnarray}
E_{qm} &=& \sum_{n=0}^\infty E_n = \sum_{n=0}^\infty \left(n + \frac{1}{2}\right)\hbar\omega\nonumber\\
&=& \frac{e^{- \frac{1}{2}\beta\hbar\omega}}{1 - e^{-\beta \hbar\omega}} \label{eqm} 
\end{eqnarray}
with $n = 0, 1, 2, \dots$, a positive integer. The average energy of a quantum harmonic oscillator is
\begin{eqnarray}
\langle \epsilon_{qm}\rangle &=& \hbar\omega \left(\langle n\rangle +\frac{1}{2}\right),\\
\langle n\rangle &=& \frac{1}{e^{\frac{\hbar\omega}{kT}} - 1}.
\end{eqnarray}
At low temperatures $kT$ is much smaller than the energy level spacings and $\langle n\rangle$ lies between 0 and 1, close to 0. Hence, $\langle \epsilon_{qm}\rangle \simeq \frac{1}{2}\hbar\omega$. 

The classical result as
\begin{equation}
Z_{cl} = \frac{1}{h}\int_{-\infty}^\infty \text{exp}\left[- \frac{\beta}{2m}p^2\right]dp \times \int_{-\infty}^\infty \text{exp}\left[- \frac{\beta}{2}m\omega^2 x^2\right]dx = \frac{1}{\beta\hbar\omega}
\end{equation}
with $\beta = \frac{1}{kT}$. 
Hence,
\begin{equation}
\langle \epsilon_{cl}\rangle = kT^2 \frac{\partial}{\partial T}\text{ln Z} = kT.
\end{equation}
The partition function of the system with only the zero-point quantum fluctuation included can therefore be written as 
\begin{equation}
Z  = e^{-(1 - \lambda)\beta \frac{1}{2}\hbar\omega} e^{-\lambda\beta \left(\beta^{-1}\text{ln} \beta\hbar\omega\right)}\label{z}  
\end{equation}
and
\begin{equation}
\text{ln} Z = -(1 - \lambda)\beta \frac{1}{2}\hbar\omega + \lambda\text{ln}\frac{1}{\beta \hbar\omega} \label{z2}
\end{equation}
This shows that an ensemble of harmonic oscillators is fully quantum mechanical if $\lambda = 0$ and fully classical if $\lambda = 1$. 

In general, $\lambda$ can vary with the volume $V$ of amphiphilic molecules as well as on additives. Aqueous solutions of surfactants show self-organization and the coexistence of water clusters with low density (LDL) and high density (HDL) as well as self-oscillations between these extremes \cite{mir}. The phenomenon can be modelled by an ensemble of harmonic oscillators described by the partition function (\ref{z}) with the parameter $\lambda$ taken as a sinusoidal function of time such as  $\lambda = |\sin \Omega t|$. Then $\lambda = 0$ at $t = n\frac{\pi}{\Omega}\, (n = 0, 1, 2, \cdots$), $\lambda = 1$ at $t =(n + \frac{1}{2})\frac{\pi}{\Omega}$ and $\lambda = \frac{1}{2}$ at $t = (n + \frac{1}{4})\frac{\pi}{\Omega}$.
Thus, the LDL and HDL clusters can be interpreted as quantum and classical ensembles of harmonic oscillators that undergo continuous oscillations. When $\lambda = \frac{1}{2}$ the system is half quantum mechanical and half classical, and 
\begin{equation}
\langle \epsilon_{qm}\rangle \simeq\langle \epsilon_{cl}\rangle,\,\,\,\,or\,\,\,\, \frac{1}{2}\hbar\omega \simeq kT.
\end{equation}

The Helmholtz free energy of the system is given by
\begin{equation}
F = -kT\text{ln}Z = (1 - \lambda)\frac{1}{2}\hbar\omega  - \lambda kT \text{ln}\frac{1}{\beta \hbar\omega}
\end{equation}
and the Gibbs free energy by
\begin{eqnarray}
G &=& F + kT\left(\frac{\partial\, \text{ln}Z}{\partial\, \text{ln}V}\right)_T\nonumber\\ 
&=&- kT\left[\text{ln} Z - \left(\frac{\partial\, \text{ln}Z}{\partial\, \text{ln}V}\right)_T\right].
\end{eqnarray}
It can be assumed that the volume of an aqueous solution almost does not change. Hence, it is justifiable to ignore the second term in this definition of $G$ which reduces to $F$. 

All thermodynamic quantities of aqueous solutions like entropy and $C_V$ can be calculated from the partition function given by
\begin{equation}
Z  = e^{-(1 - \lambda)\beta E_{qm}}  e^{-\lambda\beta \left(\beta^{-1}\text{ln} \beta\hbar\omega\right)}. 
\end{equation}
where $E_{qm}$ is given by (\ref{eqm}). Hence, the free energy $F$ is
\begin{equation}
F = (1 - \lambda)\left[\frac{1}{2}\hbar\omega + kT \text{ln} \left(1 - e^{-\hbar\omega/kT}\right)\right] - \lambda kT\text{ln}\frac{1}{\beta \hbar\omega}\label{f}
\end{equation}
and, assuming as a first approximation that $\lambda$ is independent of temperature $T$, the entropy $S$ is
\begin{eqnarray}
S &=& -\frac{\partial F}{\partial T}\nonumber\\
&=& -(1 - \lambda)\left[k \text{ln} \left(1 - e^{-\hbar\omega/kT}\right) - \frac{e^{-\hbar\omega/kT}}{1 - e^{-\hbar\omega/kT}}\frac{\hbar\omega}{T}\right]\nonumber\\ &+&  \lambda \left(k + k\text{ln}\frac{kT}{\hbar\omega}\right)\label{s} 
\end{eqnarray}
Plots of the dimensionless parameters $F/kT$ and $S/kT$ as a function of $u = \hbar\omega/kT$ are shown below.
\begin{figure}[H]
\centering
\includegraphics[scale=0.5]{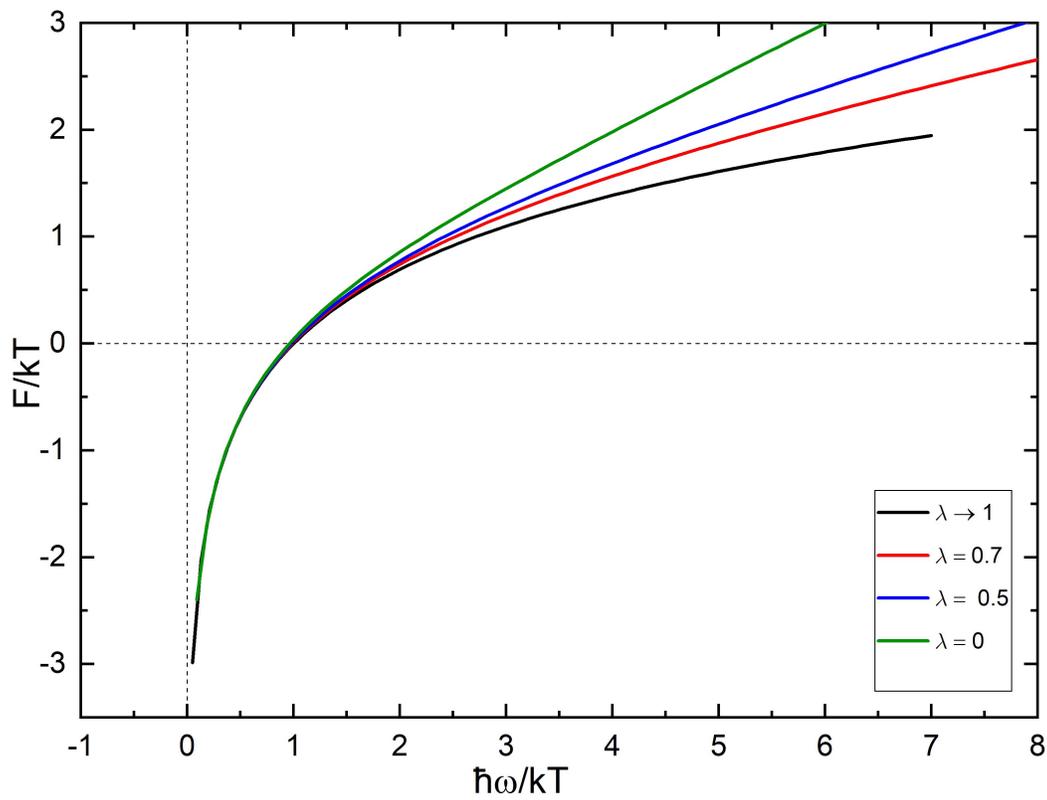}
\caption{\label{Fig. 2}{\footnotesize Plots of the ratio $F/kT$ as a function of $u = \hbar\omega/kT$ for some values of the parameter $\lambda$}}
\end{figure}

\begin{figure}[H]
\centering
\includegraphics[scale=0.5]{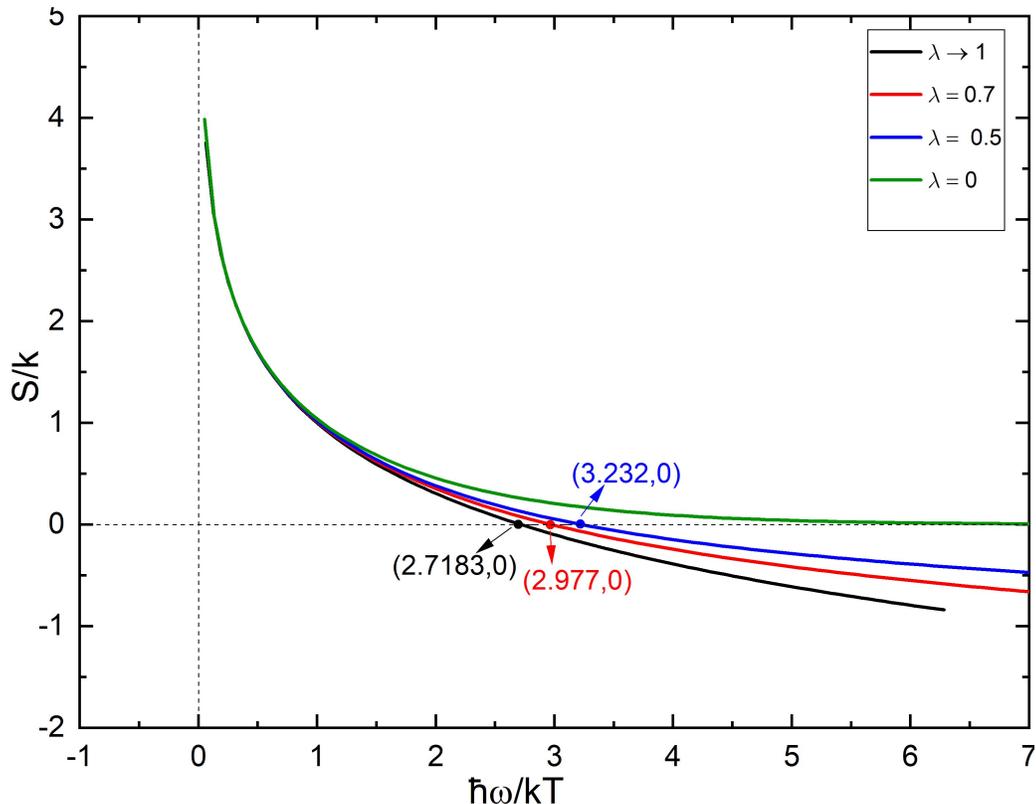}
\caption{\label{Fig. 3}{\footnotesize Plots of the ratio $S/k$ as a function of $u = \hbar\omega/kT$ for some values of the parameter $\lambda$ as per eqn (\ref{s})}}
\end{figure}
Equations (\ref{f}) and (\ref{s}) show that one has the standard results for a classical harmonic oscillator system in the limit $\lambda =1$ and the standard quantum mechanical results in the limit $\lambda= 0$. The dimensionless parameters $F/kT$ and $S/kT$ as functions of the dimensionless parameter $\hbar\omega/kT$, as expressed by equations (\ref{f}) and (\ref{s}) respectively, are given in Figures 2, 3. Figures 2 and 3 show that the quantum ($\lambda=0$) and hybrid cases ($\lambda$ fractional) converge with the classical case ($\lambda=1$) at high temperatures for a fixed value of $\lambda$, or equivalently, at low values of $\hbar\omega$ at a fixed temperature $T$.  $F/kT$ is always larger the lower the value of$\lambda$, being largest for the purely quantum mechanical case $\lambda= 0$ (green curve).  Figure 2 shows that $F$ vanishes at the point $\hbar\omega/kT = 1$ for all values of $\lambda$ as can be easily checked from equation (\ref{f}).  Furthermore, equation (\ref{s}) shows that in the classical limit ($\lambda=1$) (black curve in Fig. 3) the entropy $S$ vanishes when $\hbar\omega/kT = e = 2.7183…$, i.e. when the quantum and thermal fluctuations are approximately equal. A black arrow in Figure 3 shows this. Figure 3 also shows that the entropy $S$ vanishes at increasingly higher values of $\hbar\omega/kT$ as the system becomes more and more quantum mechanical ($\lambda \rightarrow 0$), and also that the entropy is positive for all $\lambda$ at all values of $\hbar\omega/kT\leq 2.7183$. Also, for $kT \gg \hbar\omega$, thermal fluctuations dominate and all the curves converge to the classical curve (black). These features have actually been observed in the case of micellar solutions (which can be modeled as harmonic oscillators) at T= 298 K [4]. In the fully quantum mechanical limit, the entropy is everywhere positive and, at a fixed temperature, it tends to zero as $\hbar\omega$ increases , or equivalently, for a fixed $\hbar\omega$ it tends to zero as the temperature tends to zero, consistent with the Nernst  heat theorem (the Third Law of Thermodynamics). 

\section{Concluding Remarks}
In addition to the above points, several more should be noted. As can be seen from Fig.2, the Gibbs energies of all systems reach a common minimum if $\hbar\omega/kT = 1$. Now, the minimum on the temperature dependence of the isothermal compressibility of water at 46$^\circ$ C and 25$^\circ$ C becomes clear. In the article \cite{mir3}, it is explained by the existence of LDL and HDL 1-2 nm water structures. According to the results of this article, the feature of the temperature dependence of the isothermal compressibility of water should be represented differently. Quantum fluctuations of the O-H bond of water torsion and tension oscillate in a double-well potential with proton tunneling. The energy compromise is maintained such that the energy of quantum fluctuations is equal to the energy of thermal fluctuations to maintain the minimum Gibbs energy of cavity formation in water. Thus, to create water cavities of micelles of ionic surfactants with a size of 4.2-6.2 nm (thermal fluctuations), water is compressed to a size of 0.1-0.3 nm (quantum fluctuations). The minimum temperature dependence of the change in the Gibbs energy of micelle formation of ionic surfactants shifts to 25$^\circ$ C. Therefore, the contribution of quantum fluctuations to the formation of cavities in the water for accommodating micelles increases. Thermal fluctuation (cavity) has a lower diffusion rate than 2 quantum ones, which gradually disappear with an increase in the concentration of inorganic electrolyte. The circular dichroism spectra and WAXS results confirm the competing nuclear quantum effects model for water in which its structural and dynamic properties are governed by the bias between intramolecular and intermolecular quantum contributions \cite{mir3, pat}. 

Continuing this topic, we recall that micelles played an initial role in cell evolution. With the appearance of energetically favorable water substrates with equal energies of quantum and thermal fluctuations, they evolved to cells. Water in human biological cells has a size of 4-6 nm. Therefore, the extensive parameters of the ensemble of small systems of water and other cell molecules fluctuate significantly. Because of the evolution of the cell, the condition $\hbar\omega \approx kT$ exists in it at 36.5 degrees. Therefore, the frequency of quantum fluctuations in a double-well potential should be in the terahertz range. The terahertz range of electromagnetic radiation is a promising range for non-traditional treatments, and perhaps the main treatment of a person in the future. Humans exist on the verge of equality of the energies of thermal and quantum fluctuations of water. If the energy of one of these fluctuations predominates, a person becomes unstable to infections or gets sick. 

\section{Authors' contributions}

Y. M. designed and directed the research, and contributed to the interpretation of the results. P. G. carried out applications of the interpolating Schr\"{o}dinger equation and developed the thermodynamics of harmonic oscillators for micellar solutions. The authors made equal contributions to the writing of the article.

\section{Conflicts of Interest}
The authors declare that there is no conflict of interest regarding the publication of this paper.

\section{Acknowledgement}
The authors are grateful to Anirban Mukherjee for collaborating to develop the double-well potential section, to Partha Nandi for plotting Figs 2 and 3 and Alexander Pribylov for help in preparing the manuscript. For Y. M.  the study was financially supported by the Ministry of Education and Science of the Russian Federation (g/z 2020 No. 0851-2020-0035). The study was carried out as part of the strategic academic leadership program "Priority 2030" (Agreement No. 075-15-2021-1213).

\end{document}